# Cubesats in Low Earth Orbit: Perils and Countermeasures

***Gianmario Broccia**[a]*

[a] *M.Sc. in Energy Engineering, graduated at the University of Cagliari – Department of Electrical and Electronic Engineering*

ABSTRACT

In orbit, we find a harsh environment able to damage even space-qualified components. The main threats will be listed in the following lines, one by one, also presenting some of the effects on commercial electronics. According to the literature, the most recommended materials and countermeasures will be also introduced under each "Materials and Countermeasures" paragraph.

## 1. Temperature, Vacuum and Heat Transfer

### *1.1 Hazards*

Satellites usually deal with the Low Earth Orbit, an airless, almost perfectly void environment. The vacuum at LEO is typically 1,33 · $10^{-6}$ ÷ 1,33 · $10^{-8}$ bar **(1) (2)**. The almost non-existent atmospheric inertia allows the solar radiation (which ranges between 1321 W/m² and 1412 W/m²) to directly hit any exposed surface, leading to very high temperature, depending on the reflectivity, absorptivity and heat capacity of the material itself. On the other hand, on the night side, no solar radiation is present, and the temperature might drop to several tens of degrees below zero. This mechanism leads to severe jumps in temperature and consequent thermal expansion/contraction that might result in cracks in a PCB.

The sun radiation is not the only one to be considered since the light reflected by Earth and its surface radiation are both sensible contributions. The latter, in particular, will be always present both on the day and night side, preventing any material to indefinitely decrease its temperature.

Approximately, the Earth's reflected energy is 29% so that roughly 99 W/m² are reflected into space, while the remaining 71% is emitted both by the atmosphere and the surface, resulting in about 241 W/m² emitted mostly in the infrared band **(3)**, as visible in Figure 1. Earth thus emits a black body with a temperature of 255,3 K.

* *Corresponding author.* Linkedin: https://www.linkedin.com/in/gianmario-broccia-827737165/
E-mail address: gianmario.broccia@gmail.com

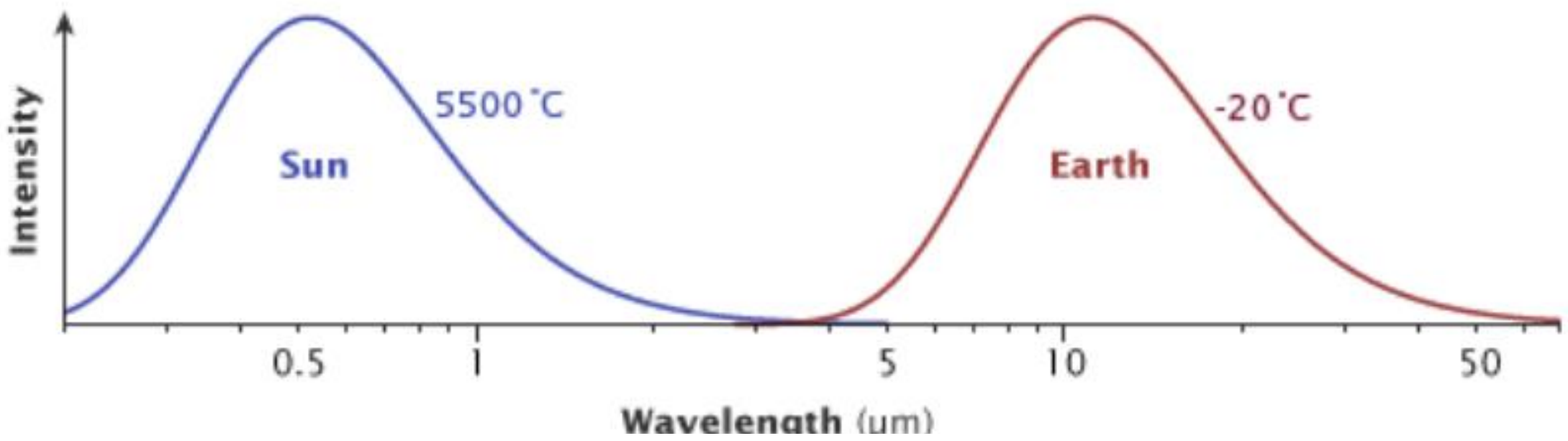

Figure 1 - Wavelengths for Solar and Earth radiation

The vacuum of space does not allow heat transfer by convection so that the only way to get rid of the exceeding heat is by radiation only or conduction (when the heat is passed to a component that needs heating). This is especially problematic for those electronics that commonly work at high temperatures, being so a heat source, since the most efficient way to dissipate the heat (convection) cannot be exploited.

*1.2 Materials and Countermeasures*

Considering that PCBs work on average between -20 and 70 °C **(4)** the quest for proper thermal control becomes the main issue. In **(5),** Chapter 7.0, a general summary of the methods commonly being used is shown.

Protections can be passive, as for the case of Multi-layer insulation which is reported not to be truly effective for small surfaces, also considering their sensibility to compression (which complicated the mounting inside a fairing or a second stage). Reflective layers, radiators, and heat pipes can be also used. Research of the Netherlands Aerospace Center proved the heat pipes, in particular, to be an interesting penalty-free solution to manage the internal heat of the satellite, moving it from a heat source to a sink **(6)**.

The operative temperature remained to be well within the tolerance of the piece and even the start with a frozen internal fluid is shown to be overtaken within 1 minute as a thermal load of 10W is applied, preventing any damage to the cooled component.

When passive methods are not enough, protections can also be active as for electric band heaters and cryocoolers. The first ones are mostly chosen to keep cold-sensible components heated, such as batteries.

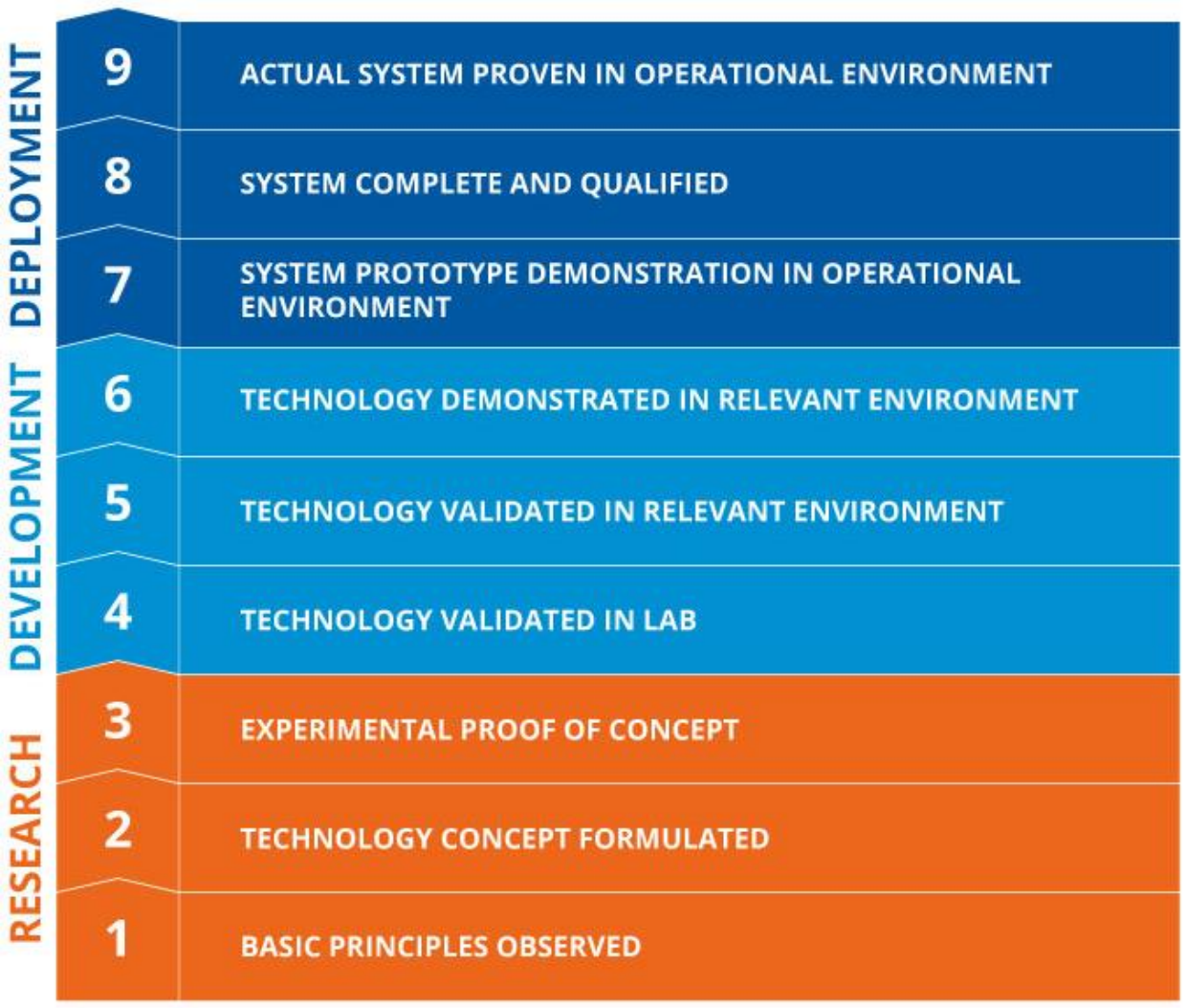

Figure 2 - Technology Readiness Levels **(7)**

On the other hand, such kinds of components imply further volume to be added and accurate control.

Cryocoolers, Fluid loops, and deployable radiators are active technologies yet to be fully qualified for regular use. Their NASA Technology Readiness

Level ranges from 4 to 7 (Figure 3).

## 2. Outgassing

### *2.1 Hazards*

An interesting description of effects on the materials is given by Boudjemai et.al. **(8).** They highlight the problem of outgassing especially for polymers and sublimation for metals with consequent formation of "whiskers" between different parts of the systems, generating the possibility of short circuits.

They also provide a useful (yet, probably approximative) mathematical relation to quantifying the rate of sublimation for a given material:

$$G = 5{,}04 \cdot 10^3 \cdot P \cdot \sqrt{\frac{M}{T}} \qquad [g/cm^3 day]$$

where:
G is the rate of sublimation $g/cm^3$day;

P is the vapor pressure of evaporating material in mmHg;

M is the Molecular weight of the material;

T is the absolute temperature in K.

As a matter of example, the Vapor Pressure of the Zinc at 393,15 K is $5{,}59546 \cdot 10^{-7}$ mmHg and so, keeping the chosen temperature, the sublimation rate would be equal to $1{,}15 \cdot 10^{-3}$ $g/cm^3$day.

The Zinc itself, along with Tin, Magnesium and Cadmium are cited by this paper among the materials to be avoided due to their sublimation rate below 200 °C. As far as concerns the use of composite material made for example from Kevlar, their attitude to the retention of humidity can trigger a marked outgassing problem and lower the structural performances.

Polymers are however not prohibited for space use, but they should be carefully tested to ensure low-outgassing properties.

### *2.2 Materials and Countermeasures*

The problem can be mitigated, especially for polymers, with a special bake-out in which the material is forced to outgas at high temperatures in a vacuum chamber. Nasa **(2)** advises a bakeout with a duration of 24h at a temperature higher than the one expected into orbit or 100°C in case this is unknown.

For PCB, Triana et.al. **(9)** recommend the use of coatings such as paraxylenes, which offer protection to an extreme temperature ranging from -200 °C to 150 °C and high tolerance to radiation. According to Plasma Ruggedized Solutions, a coating in particular is fit for use in space: the Arathane 5750-A/B **(10)**. According to what is reported on their website, this product, when cured, meets NASA outgassing properties critical for applications in outer space and high vacuum environments and is typically recommended for encapsulating modules with complicated circuitry and/or stress-sensitive components like PCBs.

With a maximum temperature of 130°C and being Mil-spec MIL-I-46058C approved (see the datasheet), this component appears to be one of the most chosen to protect the electronics of Cubesats from outgassing.

According to G. Lee, from the Fermi National Accelerator Laboratory (Batavia, Illinois) **(11)** Teflon and Kapton are good plastic materials, that can be used by virtue of their low outgassing properties and resistance to high temperatures. Also, ceramic materials should be preferred when possible.

| MATERIAL | % TML | % CVCM | CURE TIME | CURE TEMP | AT- MOS | % WVR | DATA REF | APPLICATION | MFR CODE |
|---|---|---|---|---|---|---|---|---|---|
| BLACK DELRIN LOT # S-86585-Z | 0.50 | 0.02 | | | | 0.18 | GSC28744 | MOLDING COMPOUND | DUP |
| BLACK DELRIN LOT # S-86585-Z | 0.29 | 0.00 | 24H | 125 | E-5 | 0.15 | GSC28783 | MOLDING COMPOUND | DUP |
| DELRIN 100STNC10 WHITE ACETAL | 0.78 | 0.10 | | | | 0.24 | GSC16959 | MOLD CPND | DUP |
| DELRIN 107 BLACK | 0.62 | 0.01 | | | | 0.20 | GSC13380 | MOLDING CPND | DUP |
| DELRIN 150 ACETAL 1.25 INCH DIA BLACK ROD #1 | 0.43 | 0.02 | | | | 0.20 | GSC25720 | MOLDING CMPD | DUP |
| DELRIN 150 ACETAL 1.25 INCH DIA BLACK ROD #2 | 0.47 | 0.02 | | | | 0.13 | GSC25723 | MOLDING CMPD | DUP |
| DELRIN 150 SA WHITE ACETAL | 0.28 | 0.02 | | | | 0.13 | GSC17976 | MOLD CPND | DUP |
| DELRIN 500CLNC10 WHITE ACETAL | 0.58 | 0.09 | | | | 0.14 | GSC16962 | MOLD CPND | DUP |
| DELRIN 507 BLACK ACETAL | 0.37 | 0.02 | | | | 0.14 | GSC17978 | MOLD CPND | DUP |
| DELRIN 511P, NC010 | 0.31 | 0.00 | | | | 0.15 | GSC32524 | MOLDING COUMPOUND | DUP |
| DELRIN 550 ROD - WHITE | 0.39 | 0.02 | | | | 0.10 | GSFC6953 | MOLD CPND | DUP |
| DELRIN 570-NC000 WHITE GLASS REINFORCED ACETAL | 0.33 | 0.02 | | | | 0.11 | GSC16919 | MOLD CPND | DUP |
| DELRIN D500AF BROWN MOLDING CPND | 0.30 | 0.02 | | | | 0.07 | GSC16730 | MOLD CPND | DUP |
| DELRIN II 100NC10 ACETAL WHITE | 0.34 | 0.01 | | | | 0.15 | GSC16847 | MOLD CPND | DUP |
| DELRIN II 500NC10 ACETAL WHITE | 0.28 | 0.01 | | | | 0.13 | GSC16850 | MOLD CPND | DUP |
| DELRIN II 900NC10 ACETAL WHITE | 0.29 | 0.01 | | | | 0.13 | GSC16853 | MOLD CPND | DUP |
| TK-AD DELRIN NOIR SAMPLE FROM FRANCE | 0.41 | 0.02 | | | | 0.16 | GSC28552 | PURGING CAP | ZZZ |

Figure 3 - Data on several types of Delrin **(12)**.

The abovementioned materials can be “confirmed” as low-outgassing also envisioning the NASA Online Database for Outgassing Data for Selecting Spacecraft Materials **(12)**. Here it is possible to find several materials setting a recommended upper limit for the Total Mass Loss (< 1%) and the Collected Volatile Condensable Material (< 0,1%). The first is defined as mass loss of the sample, determined from the weights before and after the 398 K exposure, expressed as a percentage; the second is defined as the difference between the weight of a clean collector and the weight of the same collector with condensed volatile materials.

Another useful material to be found inside the Nasa Database is the Delrin (Polyoxymethylene, melting point 175 °C) which presents like a resin.

## 3. Vibrations

### *3.1 Hazards*

Boudjemai et.al. **(8)**, again, provide the first hint about vibrations, which will of course depend on the launch vehicle. Any launcher passing through the atmosphere would make the flight hardware experience an intense vibratory phenomenon, which leads to the necessity of correctly devising the way it will be connected to the related stage. Under-estimating the vibrations would result in resonances capable of damaging the payload, while over-estimations would lead to sacrificing useful mass for nothing.

Mihail Petkov provides, in **(13)** a detailed description of different types of vibrations:

1. **Acoustic Vibrations** are given by fluctuation in pressure as the rocket passes through the atmosphere and by the turbulence in the surrounding air created by the gases exiting the nozzle of the rocket. This condition usually persists until the Max-Q (max dynamic pressure) is reached, going then to decrease. The vibrations are usually between 20 and 10.000 Hz.
2. **Random Vibrations**, which are accelerations manifesting within the range of 10-2000 Hz. The fairing is the main cause due to, again, the dynamic force exerted on the vehicle.
3. **Pyrotechnic Shock** associated with staging and on-orbit brief ignitions. This impulse is not usually representable with a simple function since it is transferred to the whole structure at once.

*3.2 Materials and Countermeasures*

As far as concerns the acoustic vibrations, the payload should be able to maintain the required constraints for at least 3 minutes.

For the random vibrations, the power spectral density of the random vibration should be determined first, to consequently predict the response of the payload. In this case, the requirement is represented by a first 6db/octave slope within 20-50 Hz, a constant Power Spectral Density (PSD) level dependant on the payload mass and a final -4,5db/octave slope within 500-2000 Hz. The relation defining the PSD (in $g^2$/Hz) constant level is:

$$PSD(m) = 0{,}1 \cdot \frac{m+20}{m+1}$$

Again, the duration of the test should be 3 minutes.

Finally, for the Pyrotechnic Shock, is the first slope up to 1500 Hz of 9 dB/octave and a constant part afterward. This test is supposed to be applied for each of the three axes.

Apart from the prescriptions for a correct setup of the payload in the launcher fairing contained in the ECSS-E-HB-32-26A (Spacecraft mechanical loads analysis handbook), there is no particular attention to the use of specific materials for PCBs, since the vibrations can be mitigated by using external mounts.

## 4. Radiations

*4.1 Environment*

A handbook created by Texas Instruments **(14)** (especially for the effects on electronics) and a Ph.D. Thesis by J. Rushton **(15)** are the main sources from which this paragraph comes from. Any further reference is reported within the related statement.

First, radiations are divided into three main categories:

- **Solar Radiation**, including lower-energy photons, plasma and occasional solar flares;
- **Cosmic Rays**, mostly composed of high-energy (~ 1 GeV at roughly 1 particle/$m^2$s) protons coming from all directions with same intensity;
- **Radiation Belts**, which are represented by zones in which energetic particles accumulate due to the terrestrial magnetic field, forming hazardous toroidal “bands” all around the planet.

*4.1.1 Solar Radiation*

Solar Radiation is divided into three components: Solar Wind, Solar Flares and Coronal Mass Ejections (CMEs).

Despite the constant and intense flux, the Solar Wind is the least hazardous component, since it is composed of low energy protons (<1keV) and electrons (<100eV) ejected at 400-800 km/s.

Solar Flares generate a burst of radiations in the whole spectrum and all directions, involving energies several orders of magnitude higher than solar winds. Along with the particle energy, another danger is given by the speed of those bursts which travel at light speed and are close to being unpredictable.

Unlike Solar Flares, Coronal Mass Ejections are represented by an ejection of coronal material ($10^9$-$10^{13}$ kg at speeds ranging from 20 to 3000 km/s) which reaches the Earth in approximately 3 days, carrying high-energy electrons, protons, and heavier ions and the associated flux can reach 500.000 particles/cm$^2$s.

Solar Flares and CMEs are commonly associated with Solar Energetic Particles (SEP), which energy ranges from tens of MeV to GeV with a maximum ever observed of 20 GeV in a Solar Flare.

### *4.1.2 Cosmic Rays*

Cosmic Particles under 50 MeV are not able to penetrate the termination shock, a giant edge where cosmic rays and solar wind are in balance leading to the blockage of 75% of the incoming particles. Only particles with energy around 1 GeV can go through the heliosphere and their related flux decreases as their energy increase: the higher is the energy, the rarer is the particle.

### *4.1.3 Radiation Belts and SAA*

Radiation belts, also mentioned as Van Allen belts, are usually comprised between 1200 and 60000 km and can be represented as a toroidal tilted of 11° with respect to the rotation axis. For this reason, the inner belt converges 200-800 km above Brazil, forming the South Atlantic Anomaly (SAA) between 0 and -50 degrees of latitude and -90 and 40 degrees of longitude. In the inner belt (1200–6000 km) are electrons with energies mostly around 1-5 MeV and protons around 10 MeV, while the outer belt mainly hosts electrons with 10-100 MeV.

Despite the presence of low-energy particles, their density is hazardous in case of electronics undergo a prolonged exposure. Especially in the SAA, the particle flux is higher than other zones at the same height, leading to a potentially dangerous zone in the LEO (depending on the orbit). The ISS, for example, receives the majority of the radiation dose while passing through the SAA.

Suparta et.al. **(16)** provides an interesting deepening about this particular topic, starting from data retrieved by the National Oceanic and Atmospheric Administration (NOAA) 15 satellite (804 x 821 km polar orbit) in a time horizon ranging from 8 September to 28 October 2003, to create a model fitting the electron distributions gathered by the satellite itself.

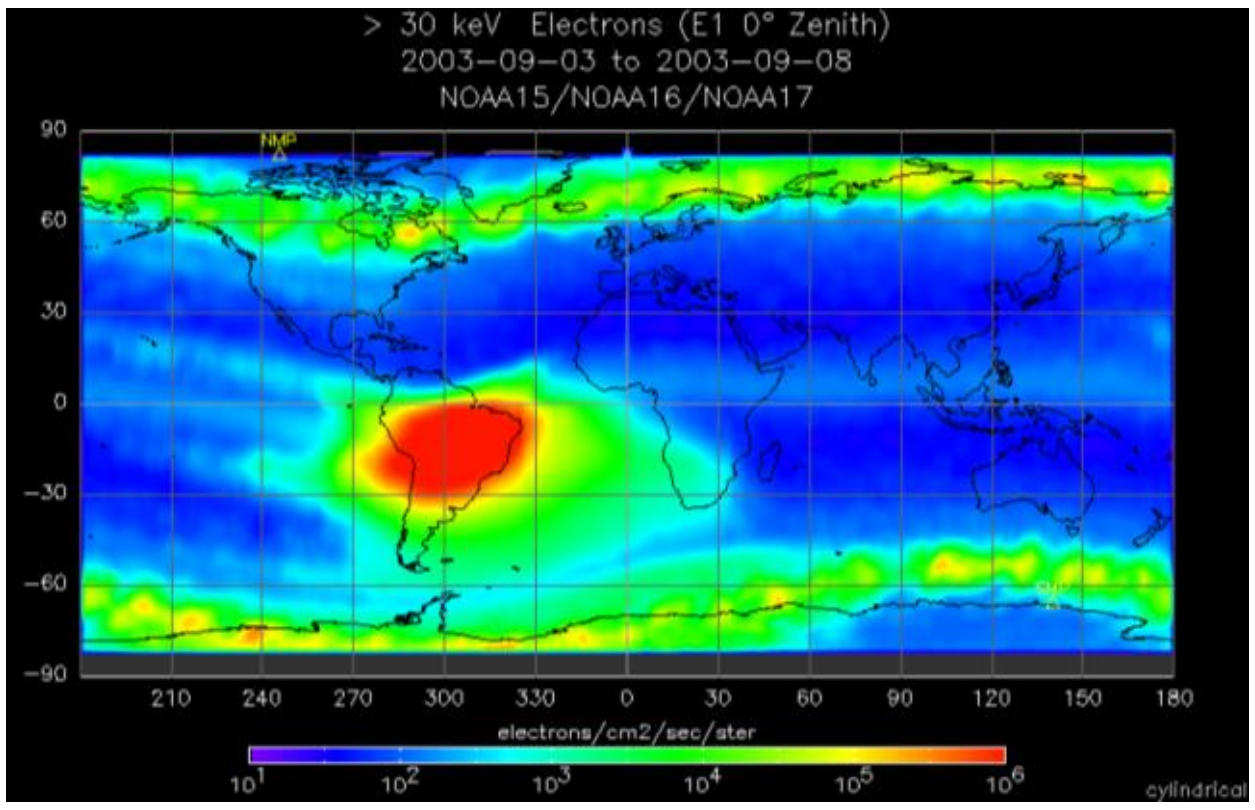

Figure 4 - NOAA image distribution of >30 keV electron from 23-28 Oct

In the image is a distribution of the electrons above 30 keV obtained from NOAA 15 data, which is indeed characterized by a marked concentration within the SAA.

Another study by Heirtzler **(17)** interestingly recalls some measurements of the dose rate based on data from the MIR space station (March 2-11 1995) and the Skylab (December 1973 – January 1974) and valid for a 400 km altitude.

It is clear how the dose in nGy/minute (100 rad = 1 Gray) is up to 7 times higher in the SAA for the Skylab and almost 3 times higher for the MIR respected the rest of the environment. Equatorial or high inclination orbits are not safe from the effects of this zone and a satellite that is forced to operate in such orbits will have to encounter the SAA multiple times a day.

From Heirtzler, again, is another interesting piece of information about the SEU suffered by the Topex/Poseidon satellite between 1992 and 1998. Despite the far high-altitude of 1340 km, not fit for smallsat, it is worth noting how the majority of SEUs occurred above the SAA (fig. 5), within the region where the geomagnetic field has its lowest value of 20.000 nT.

*4.2 Dose Effects*

This phenomenon is related to slow and increasing deviance from the wanted working conditions due to a recurrent radiation exposure that eventually leads to a fatal point of no return.

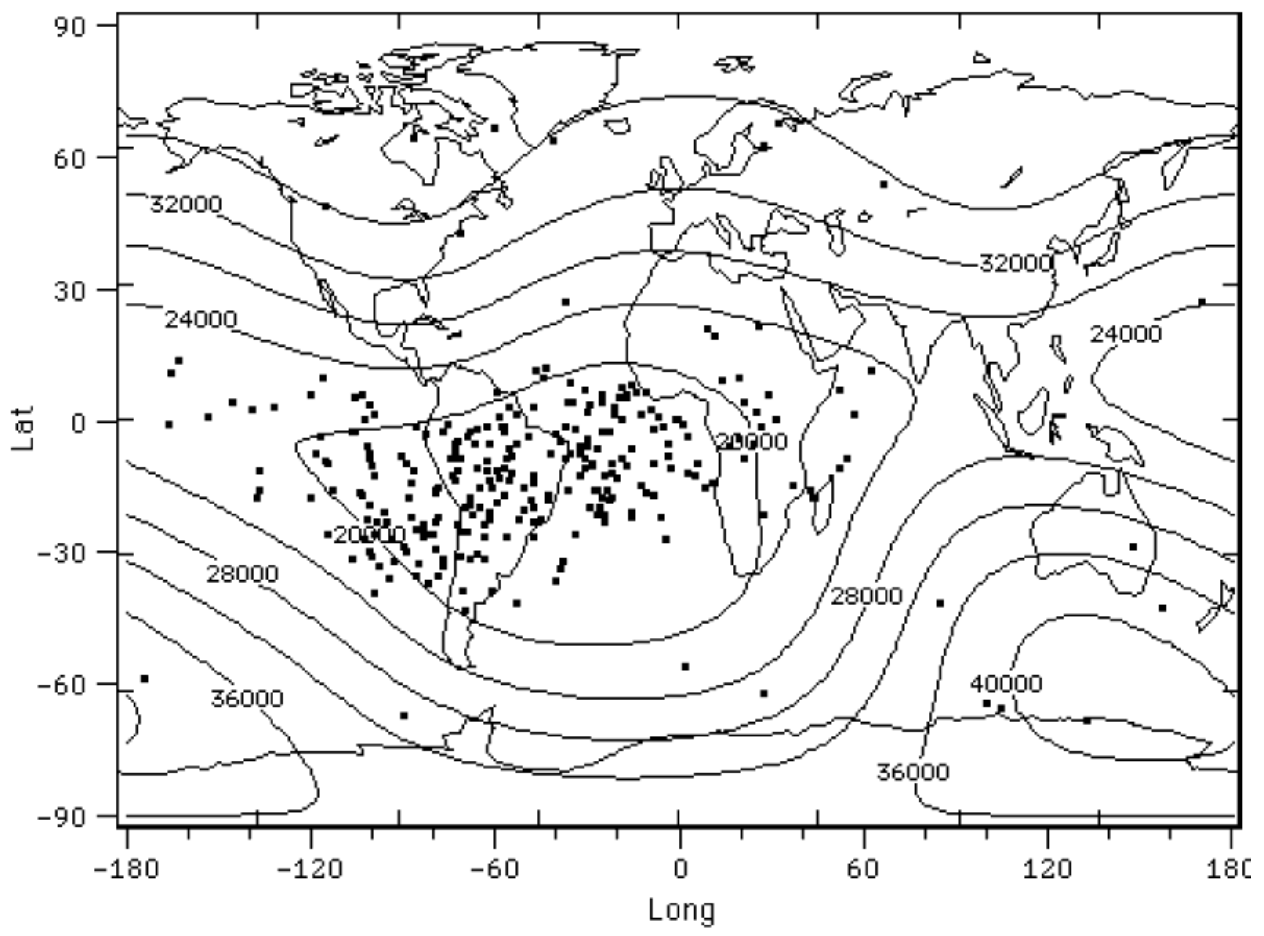

Figure 5 - Topex SEUs above the SAA

Two are the possible effects: the Total Ionizing Dose and the Displacement Damage. As reported in some material from the Massachusetts Institute of Technology (MIT), the radiation dose can be defined as the amount of energy from ionizing radiation deposited in a mass of some material **(18)** according to:

$$D = \Delta\varepsilon/\Delta m$$

where Δε is the mean energy transferred by the radiation to a mass Δm.

*4.2.1 Total Ionizing Dose (TID)*

The TID represents the energy absorbed per unit of mass when exposed to ionizing radiation and the commonly used unit is the Rad (1 Gray = 100 Rad). The accumulation of charges can lead to a series of problems but, as long as the material is a conductor or semiconductor, the charges do not accumulate and are dissipated by both recombination and diffusion/drift. When it comes to insulants, the band gaps are consistently wider, and the carriers have very low mobility. As a matter of example, Silicon Dioxide (a common insulator in semiconductor devices, especially for gates) is rapidly degraded by exposure to radiation.

A presentation by Marc Poizat (ESA) **(19)** lists a series of standards that can be consulted for a proper TID testing which, Poizat reminds, are destructive and cannot be done on components bound to be sent in space:

- ESCC 22900 - Total Dose Steady-State Irradiation Test Method. Issue 5 was released in 2016 **(20)**.
- MIL-STD883G Method 1019.7 - Ionizing Radiation (Total Dose) Test Procedure **(21)**.
- ASTMF1892-06 - Standard Guide for Ionizing Radiation (Total Dose) Effects Testing of Semiconductor **(22)**.

A material undergoes ionization damages when hit by a high-energy photon or charged particles and electron-hole pairs (ehp) are created. The kinetic energy of the incident particle is partially dissipated by creation of the pairs themselves and their density is dependent on the ionization energy of the hit material and its density. A paper by Barnaby et.al. gives an analytical representation of such dependence:

$$k_g = \frac{100 \cdot \rho}{(1.6 \cdot 10^{-12}) \cdot E_p}$$

where $k_g$ is the ehp density in ehp/cm$^3$rad and $E_p$ is the ionization energy of the material **(23)**.

*4.2.2 Displacement Damage (DD)*

It occurs with radiation-induced neutron dose/proton dose (ND/PD) related to the accumulation of physical damage and respects the TID (which involved the surface), the DD involves the whole volume. The damages are owed to a ballistic intervention of the striking ion (projectile) which displaces one atom from its position in the crystal structure of a semiconductor. The defects generate asymmetries that lead to a great change in the optical, thermal and electrical features of the zone around the defect itself. Examples of effects are electron trapping that lowers the charge transfer in CCD sensors and the increasing of electron scattering which decreases their mobility **(24)**.

With the accumulation of those, such deviations become visible at a macroscopic level, manifesting with a drop in performance and failures. The particles associated with such phenomenon are energetic electrons, proton, and neutrons but also gamma rays and X-rays (millions of eV) can produce electrons with high kinetic energy by interacting with the target material.

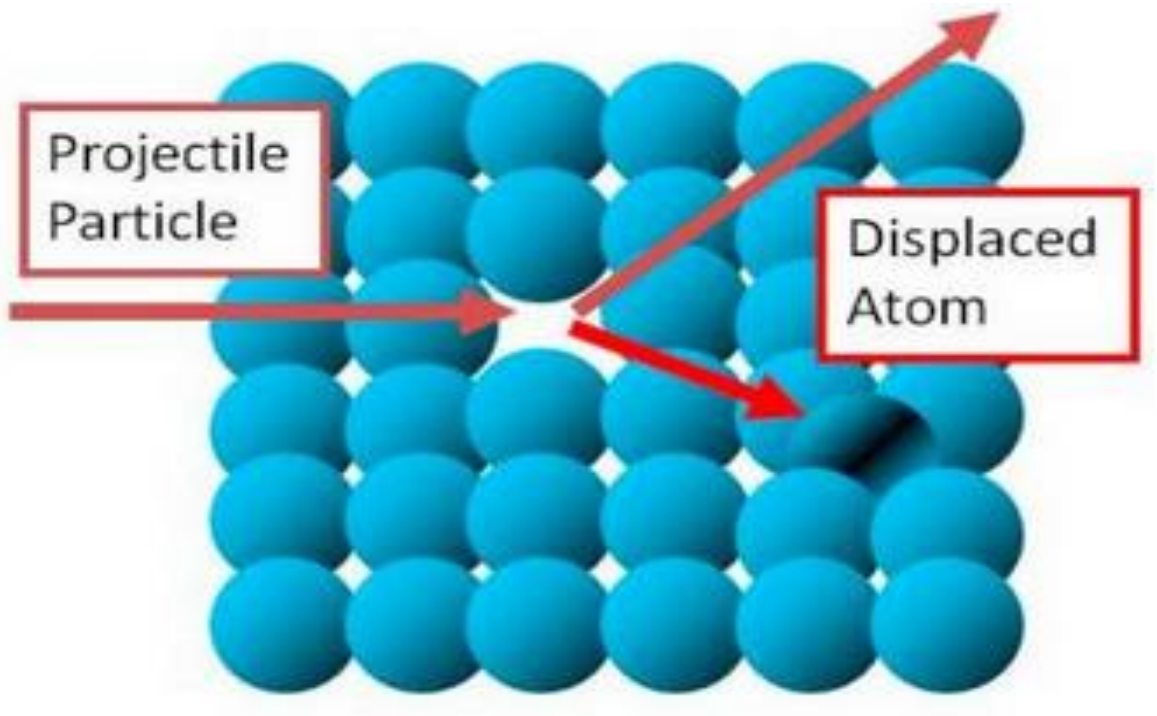

Figure 6 - Displacement Damage principle

*4.3 Effects on the Electronics: Single Event Effects*

The following part is mainly based on Chapters 3 and 4 Texas Instruments Handbook. Again, any additional citation is promptly highlighted.

Equatorial LEO is presented as the less effective in terms of radiation effects since the magnetic field expresses its greatest protection against external radiation sources, but spacecrafts with a highly inclined orbit will need to minimize the exposition during an eventual passage through the SAA. The danger for electronics is represented by Single-Event Effects (SEE) and Dose Effects.

The European Aviation Safety Agency (EASA) defines the SEEs according to which "[..] SEEs occur when atmospheric radiation, comprising high energy particles, collide with specific locations on semiconductor devices contained in aircraft systems. Memory devices, microprocessors, and Field Programmable Gate Arrays are most sensitive to SEE. Some examples of these types of effects are Single Event Upsets (SEU), Multiple Bit Upset (MBU), Single Event Gate Rupture (SEGR), and Single Event Burn-out (SEB). However, SEU and MBU are the two single effects that present the largest potential threat to aircraft systems. […]" **(25)**.

Among SEEs several are the possibilities among non-destructive and destructive events owed to the interaction with a single particle so that subcategories can be defined.

*4.3.1 Single Event Transient (SET)*

When a single energetic ion encounters an electronic device, reaching the semiconductor substrate, it generates an excess of charge carriers that possibly do not recombine rapidly. Depending on the hit point, the excess of charges can pass to sensitive components. This will be called Single Event Transient (SET).

The SET manifests as a transient voltage pulse starting from the hit node that propagates to the device output, where it appears as the same voltage transient, an amplified version of this transient, or a change in the logical output. SETs were first identified following an in-flight anomaly in the Topex/Poseidon spacecraft already presented in 4.1.3 **(26)**.

*4.3.2 Single Event Upsets (SEUs)*

It occurs when a node of a digital storage component is hit. This leads to errors in data state and, although not dangerous by itself, it may lead to major failures if new data are not overwritten, which implies that the error must be detected first.

A report by INTEL describes the dynamics of how alpha particles and neutrons lead to an SEU. In the first case, a strike produces a path (to be intended as a "wake") along which ionized charges (electrons and holes) form and recombine, depending on the given layer. However, when this path arrives down to the depletion region underneath a drain-gate-source region the electrons it creates can be quickly attracted to a higher voltage NMOS drain diffusion, which sometimes results in the change of state of a storage element. Similarly, for PMOS transistors, the holes can be quickly attracted to a lower voltage PMOS diffusion. In the case of a neutron, despite not being charged, at high speed can however reach the depletion region causing the same ionization path as the first case. Again, this might result in the migration of electrons to a higher voltage or holes to a lower voltage, causing a change of state in a digital storage **(27)**.

Koshiishi et.al. conducted interesting research about the Japanese satellite "Tsubasa" which investigated the effect of space radiation in Geostationary orbit (GEO) and, in particular, the effect on commercial components, between 2002 and 2003. By comparing the measured proton dose with the single event upset count on test samples, the conclusion was that an adequate aluminum shield should have been about 1cm, which might be feasible for large satellites but not for CubeSats. The spatial distribution of protons above tens of MeV was found consistent with the SEU count **(28)**.

*4.3.3 Single Events Functional Interrupts (SEFIs)*

According to the Joint Electron Device Engineering Council (JEDEC), a SEFI is a soft error that causes the component to reset, or otherwise malfunction in a detectable way **(29)**.

According to Texas Instruments, a SEFI occurs when an SEU changes the state (flips) of a bit in a critical system register, starting an unwanted function of the system (such as a total reset for a self-test sequence). Respect a generic SEU, which fatal ending depends on the algorithm, SEFI surely lead to a malfunction.

The only possibility to counter the problem is to power down the device and the eventuality characterizes a “hard SEFI”. There is also the possibility of a so-called “soft SEFI” when the memory is not interested in the SEU and the data can be overwritten without resetting the memory itself. An example of mitigation is offered by Austin et.al. which proposes “Watch Dog Timers” to be associated with microcontrollers which, in turn, send back a pulse as “proof” of regular functioning. If those are hit and stop sending proper pulses, the timer sends a reset signal to a switch which resets the microcontroller thanks to an SEU-immune memory, hopefully clearing the error **(30)**.

*4.3.4 Single Events Latch-ups (SELs)*

It occurs with the generation of a low-impedance connection between power and ground which remains in place even after the cause ends. The resulting high current remains until the component fails or a power cycle is done, and this effect appears to be more present in CMOS devices and it is a particularly feared problem. Becker et.al. investigated CMOS sensors by subjecting 6 devices to at least 20 latch-ups by exposition to Californium-252 at nominal supply voltages and operating current. All the devices eventually suffered a catastrophic failure but three of them also showed what the authors define as “latent damage” and is structural damage that does not reveal by failures or observable change in behavior. This possibility is indeed insidious since the device might be mistakenly considered still safe **(31)**.

SEL sensitivity is also determined by the substrate and well doping, operating voltage, and ambient temperature. The lower the substrate and well doping, the higher the resistance, and the less charge required to initiate the forward biased condition. High temperatures lower the voltage required to trigger the SEL.

*4.3.5 Single Event Gate Ruptures and Single Event Burnouts (SEGRs/SEBs)*

This kind of event is probably the most catastrophic possible since it involves the partial or total melting of a component following a marked drop in the operating voltage. This event usually strikes Metal Oxide Semiconductor Field Effect Transistors (MOSFETs) or Bipolar Junction Transistors (BJT).

As for a guideline edited by NASA in 2008, the SEGR is an event that causes the gate not to be able to regulate the current flow between the source and the drain by damaging the gate insulator (SiO2). The gate-to-drain current suddenly increases following the irradiation. Interestingly, the report highlights that SEGRs do not depend on the temperature and are most likely to occur when the ion hits with a direction perpendicular to the vertical axis of the device.

SEBs produce the same increase in the current flow rather owed to a direct short between source and drain and are dependent on the temperature and likely to occur especially at lower ones **(32)**.

To lower the possibility of a SEGR, one approach is represented by operating a device below its normal operating limits hoping to increase its lifespan. Derating factors can be found in a specific Nasa report and are intended to lower the electrical and thermal stresses, and thus decrease the rate of degradation of the device **(33)**. This practice is also reported, proposed, and further explained in a Ph.D. Thesis by Jean-Marie Lauenstein **(34)**.

### *4.4 Materials and Protections*

#### *4.4.1 Solutions according to Texas Instruments*

Chapter 6 of **(14)** reports several two methods for radiation hardening of commercial electronics: radiation hardening by process and radiation hardening by design.

In the first case, an action to be taken can be the medication of the silicon substrate with a highly doped material which, by virtue of lower resistance, is supposed to reduce the sensitivity to single event latchups. This choice is reported to be more effective with flat devices, such as CMOS.

In the second case, only physical and spatial features are modified by:

- **Component Configuration**. As a matter of example, increasing the width of a transistor would result in more current and the ability to better compensate any charge excess given by a radiation event.
- **Component Layout**. Device-level hardening implies modifications to the component design whereas methods at the circuit level are much more frequently employed and intended to increase the drive and create drive redundancy. In this case, the sensitivity to SEL is improved at the circuit level by, for example, attaching multiple drive transistors to maintain the data state of a node. It is reported that, typically, two transistors are assigned to the same node to increase redundancy.
- **Circuit Redundancy**. Digital memories are very sensitive to radiations both for the necessity of low power and high density. In this case, the solution is to employ additional circuitry to detect any bit error. Additional bits are used to store “double data”, to ensure (when data are read) the possibility of detecting any discrepancy.

The ultimate form of system-level redundancy is the choice of double identical processor cores executing the same code at the same time. This is expensive in both area and power because the same computation and instruction flow runs on each redundant core.

#### **4.4.2 Solutions according to other studies**

Another possibility is the use of proper shielding (aluminum, tantalum, or tungsten) to mitigate the Total Ionizing Dose (TID) and possibly prevent SEUs and worse. An example is given by Nasa’s Research in which a Z-grade material has been proposed as an experiment for the Shield-1 Cubesat in Figure 8 **(35)**. They test an Al/Ta vault shield compared to Al and Ta shields alone, using the program SPENVIS, against 4-6,5 MeV electrons.

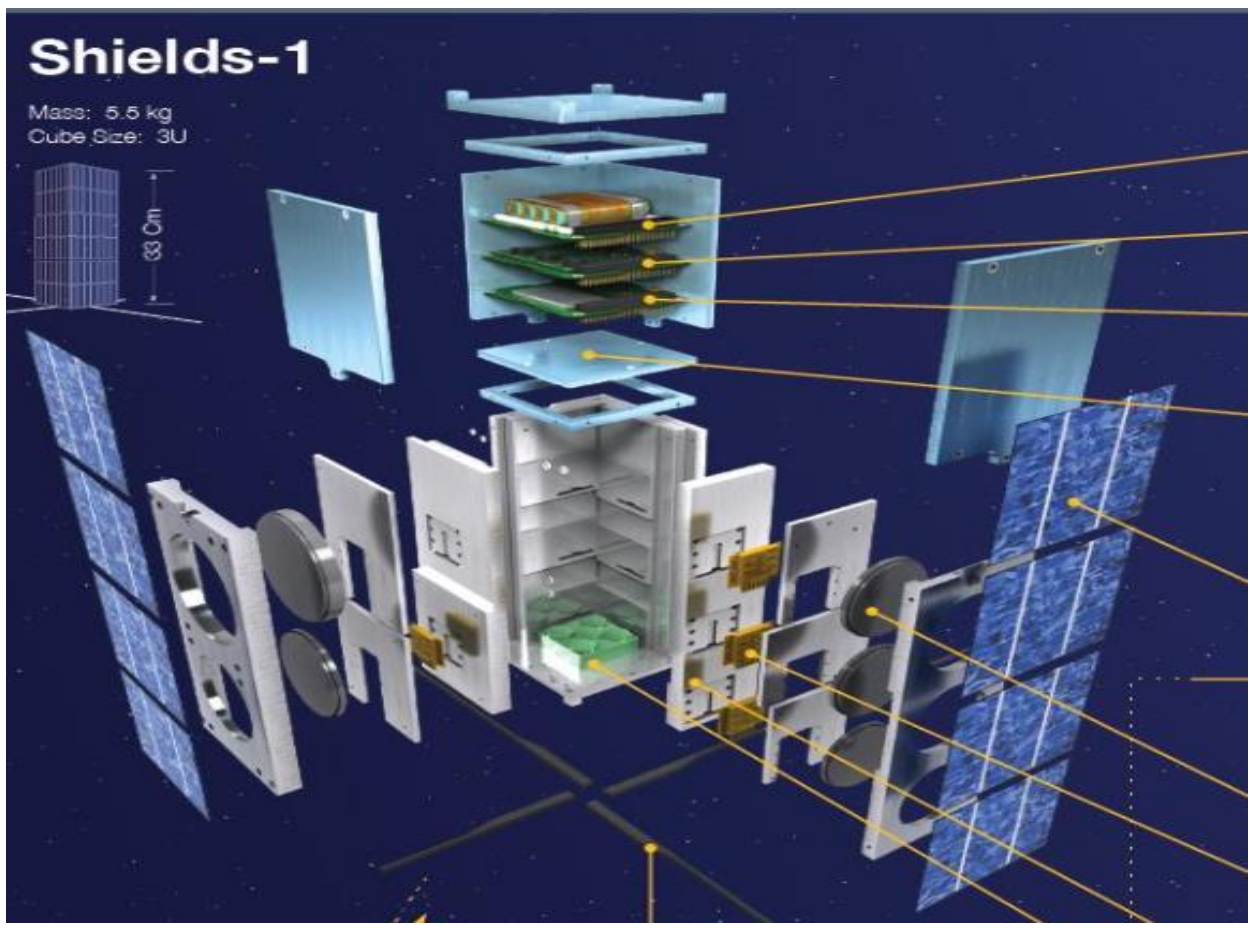

Figure 7 - Shield-1 Nasa Cubesat

As for the electron case, for areal densities above 1,7 g/cm$^2$ appears clear how the Al/Ta greatly reduces the ID, almost reaching 0 rads from 2 g/cm$^2$ on. In the proton case, the Al/Ta behaves similarly to the Al and far better than the Ta, approaching a zero dose at 3 g/cm$^2$.

The near-complete elimination of electron radiation at areal densities greater than 2 g/cm$^2$ reduces the chance of internal charging effects on electronic boards inside the CubeSat that causes anomalies. According to the paper, it seems that the thickness for the Al/Ta is 0,5 cm, while the Al is 1,1 cm, which represents of course a useful mass and volume saving.

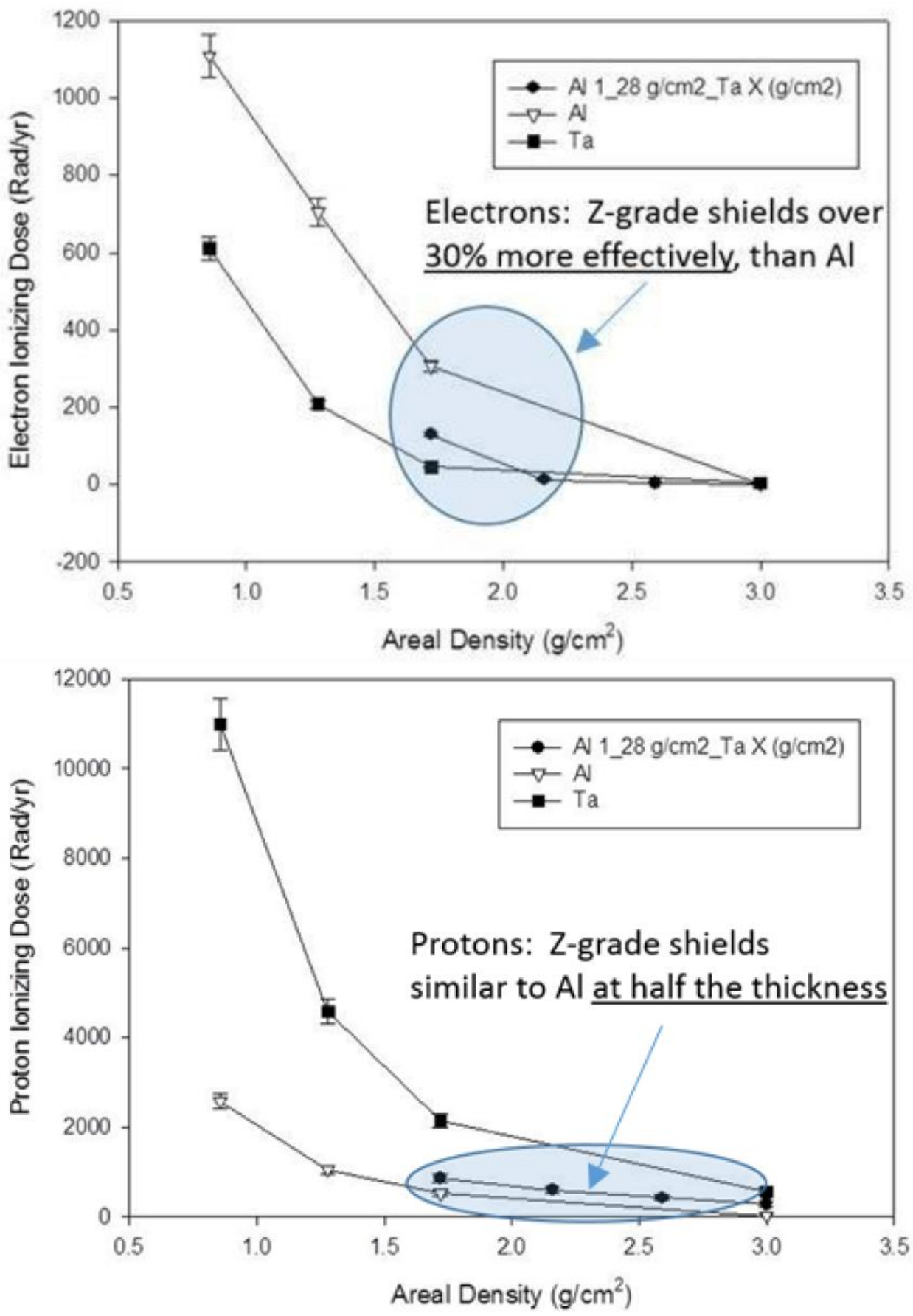

Figure 8 - Electron and Proton Ionizing Dose (35)

Also, Sinclair and Dyer **(36)** propose "Careful COTS" approach. Rather than buying space-qualified components or buying COTS to be flown "black box", a middle way is to be preferred.

According to the authors, respect the 100-1000 krad range of survival for space-qualified components, commercial components are instead tested for 30 krad. Once the components are identified and chosen the idea is to run a burn-in test on the whole assembly (CubeSat) to prevent the risk of infant mortality after the launch.

To understand the environment in which the components will work, the authors provide a modeling example for an estimation of the TID over a tot-year mission using an ESA tool called SPENVIS **(37)**. This tool allows to model mission duration, environmental conditions given by the orbit, and the shielding of the satellite, providing then the TID required to qualify or not the component under examination.

They also provide some best practices about COTS and indications about radiation testing.

## 5. Atomic Oxygen

### *5.1 Hazards*

A report by the ESA-ESTEC is useful to give a first hint about the nature of the AO **(38)**. Its formation occurs in Low Earth Orbit when Molecular Oxygen in the upper atmosphere is hit by sunlight (photodissociation) at wavelengths less than 243 nm. The low pressure prevents the oxygen from recombining and the result is that AO dominates the atmosphere from 180 to 650 km. Despite the atmosphere rotates with the Earth itself, the orbital velocity of a spacecraft is far higher resulting in an erosion phenomenon given by continuous impacts of Oxygen Particles with a mean energy of about 5 eV **(38)**.

With reference to the Materials International Space Station Experiment (MISSE), a report by Banks and Groh **(39)** summarized data and results from the missions MISSE 2, 3, 4, 5, 6, 7 and 8, offering an interesting panoramic on how the materials sent into orbit reacted. Confirming what is reported in another precedent Nasa Report **(40)**, Banks and Groh underline that even protected material can suffer from the interaction with AO, due to imperfections in the external layer which may lead the AO to penetrate and remain trapped between the material to be protected and the protection itself (Figure 10 is an example).

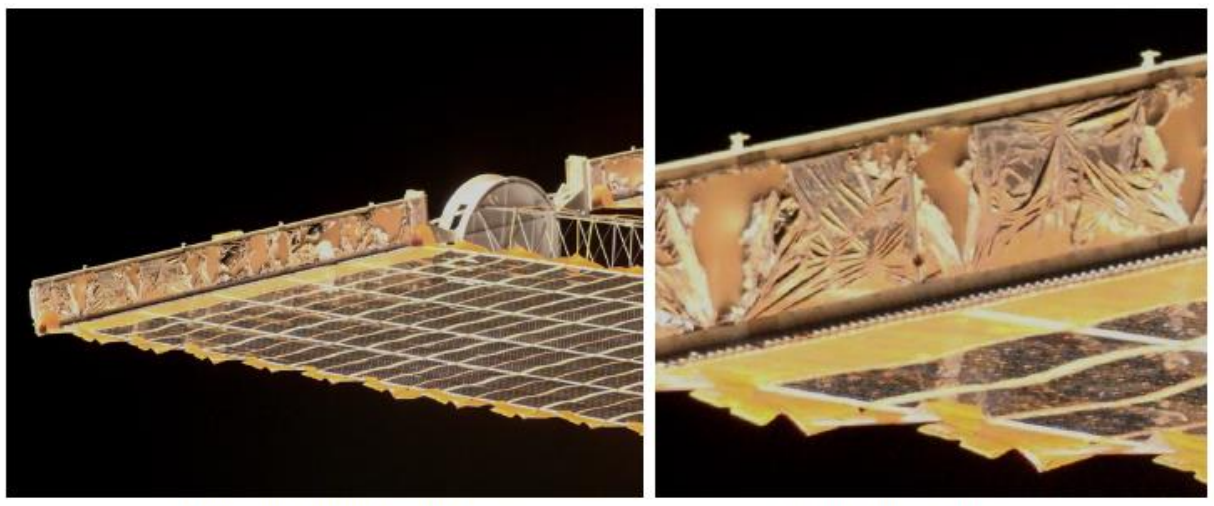

a. ISS solar array blanket box. b. Close-up of damage.

Figure 9 - Effect of AO in the Aluminized Kapton cover of the arrays of the ISS after 1 year

They also provide a basic, yet common, analytic formulation useful to calculate the Erosion Yield ($cm^3$/atom), which represents the volume lost by each AO particle strike:

$$E_y = \frac{\Delta M_s}{A_s \rho_s F}$$

where:

$E_y$ is the erosion yield of the sample in $cm^3$/atom;

$\Delta M_s$ is the mass loss of the sample (g);

$A_s$ is the surface area of the sample exposed to the AO ($cm^2$);

$\rho_s$ is the density of the sample ($g/cm^3$);

F is the low earth orbit AO fluence ($atoms/cm^2$).

Surely interesting is the fact that the Erosion Yield is directly dependant on the AO fluence itself and it is not an intrinsic property of the material itself.

The MISSE 2-8 exploited sever Kapton H samples used to determine the AO fluence with an inverse formula of the equation introduced before, known the erosion yield ($3x10^{-24}$ $cm^3$/atom). Among the materials analyzed were Teflon FEP, PTFE, White Tedlar (by DuPont), High-Temperature Polyimide (PMR-15), Clear Polyimide (CP1™), Upilex-S, Kapton HN, Mylar, and Polybenzimidazole (PBI).

*5.2 Materials and Protections*

The Erosion Yield (EY) for unprotected materials ranged from $3{,}81x10^{-27}$ $cm^3$/atom ($4{,}62x10^{21}$ $atoms/cm^2$ fluence) for the DC 93-500 Silicone **(41)** on MISSE 8 to $9{,}14x10^{-24}$ $cm^3$/atom ($8{,}43x10^{21}$ $atoms/cm^2$ fluence) on MISSE 2. The silicone has an excellent EY, considering that the general average is $10^{-24}$ $cm^3$/atom for the Kapton H, but it happens to turn into a glassy silicate which is further attacked by the AO over time. It might be still good for a short-duration mission.

The white Tedlar in particular has been found to have a decreasing EY with the increasing AO fluence, which is indeed reassuring. Polymers and other oxidizable materials, instead, can be eroded because of reaction with atomic oxygen (see table in appendix A, reporting a mix of the presented materials).

Ablative coatings are commonly used to prevent the AO from eroding the underlying layer. Such kinds of coating are usually represented by metal oxides which cannot be oxidized any further. Silicon dioxide has successfully been used on the ISS with a $1{,}3x10^{-7}$m layer by sputter deposition **(40)**.

Another study by Banks et. al. **(42)** strengthens the idea that polymers with inorganic material are less bound to be eroded by the AO thanks to the generation of non-volatile ashes that contribute to shielding the underlying material. As a matter of example, they compare in particular the yield erosion (cubic centimeters loss for each AO atom strike) of white Tedlar (0,101 · $10^{-24}$ $cm^3$/atom) with respect to the clear Tedlar (3,19 · $10^{-24}$ $cm^3$/atom). The first Tedlar basically contains a supplementary pigment made out of Titanium Dioxide.

## 6. Plasma and Spacecraft Charging

### *6.1 Environment and Hazards*

In Low Earth Orbit the gas composing the atmosphere is subject to solar radiation and experience an ionization, turning into plasma. Plasma is defined as gases with an equal number of positive and negative charged particles and thus with neutral overall charge, and yet able to conduct currents **(43)**. Their recombination, as well as Atomic Oxygen, is unlikely due to the low-pressure environment **(44)**.

Carlos Calle, from the NASA Kennedy Space Center **(45)** explains how a spacecraft moving (at about 8km/s) into an ionospheric plasma is subsonic respect electrons thermal velocities (200km/s) and supersonic respect positive ions thermal velocities (1 km/s). This leads to the formation of a wake in which the ion density is lower since ions can only impact the face normal to the satellite forward motion. Electrons can instead hit every surface and repel each other within the wake so that the wake itself lacks both electrons and ions.

A study by Fajardo et.al. **(46)** who developed a 23,5 kg smallsat called "Ten-Koh", provides little data about the plasma environment. This satellite was launched in 2018 as a secondary payload of the GOSAT-2 by a JAXA's HII-A F40 rocket. One of its goals was to research the effects of Low Earth Orbit environment, including the plasma, at an altitude of 593 x 615 km.

Carrying two Langmuir probes, the general idea was to study the formation of the plasma sheath around the satellite and possibly its variation along the orbit. The plasma density was found to range from 1.0409 × $10^4$ to 2.0248 × $10^5$ particles/$cm^3$ with an electron temperature of 2407.8 K (equivalent to 0.2 eV). Those values do not seem to be fully in line with the NASA-HDBK-4006A ver.2018 **(47)** which instead provides a table for different "zones" of the ionosphere (Fig.11).

| Region | Nominal Height of Peak Density (km) | Plasma Density at Noon ($cm^{-3}$) | Plasma Density at Midnight ($cm^{-3}$) |
|---|---|---|---|
| D | 90 | ~1.5 x $10^4$ | Vanishes |
| E | 110 | ~1.5 x $10^5$ | ~1 x $10^4$ |
| F1 | 200 | ~2.5 x $10^5$ | Vanishes |
| F2 | 300 | ~1.0 x $10^6$ | ~1.0 x $10^5$ |

Figure 10 - Nominal Properties of Ionospheric Layers

According to the same report, above 300 km the electron density in the upper layer (F2) decreases monotonically along several earth radii. Electron and ion energies are instead in line between the aforementioned report and Fajardo et.al., being quantified in 0.1 – 0.2 eV, reaching thousands of eV at geosynchronous orbits (above 40000km).

The same NASA-HDBK-4006A ver.2018 also provide a thorough analytical representation of plasma interactions, referring first to the Poisson Equation, which governs the potential distributions that drive the charge movement:

$$\nabla^2\varphi = -\rho/\varepsilon_0$$

where respectively are the potential, the charge density, and the permittivity in a vacuum. Electrons are lighter than ions, thus easier to be collected. This leads the surface involved to change its potential with respect to the surrounding plasma so that the overall net current is zero, i.e in equilibrium:

$$I_{ion} + I_{electron} + I_{photoelectrons} + I_{secondary\ electrons} + I_{other\ sources} = 0$$

If the acquired potential is lower than -100 V respect the surroundings, the involved surface will be subject to arcing (i.e. a sudden charge transfer between points at different voltage) lasting micro or milliseconds.

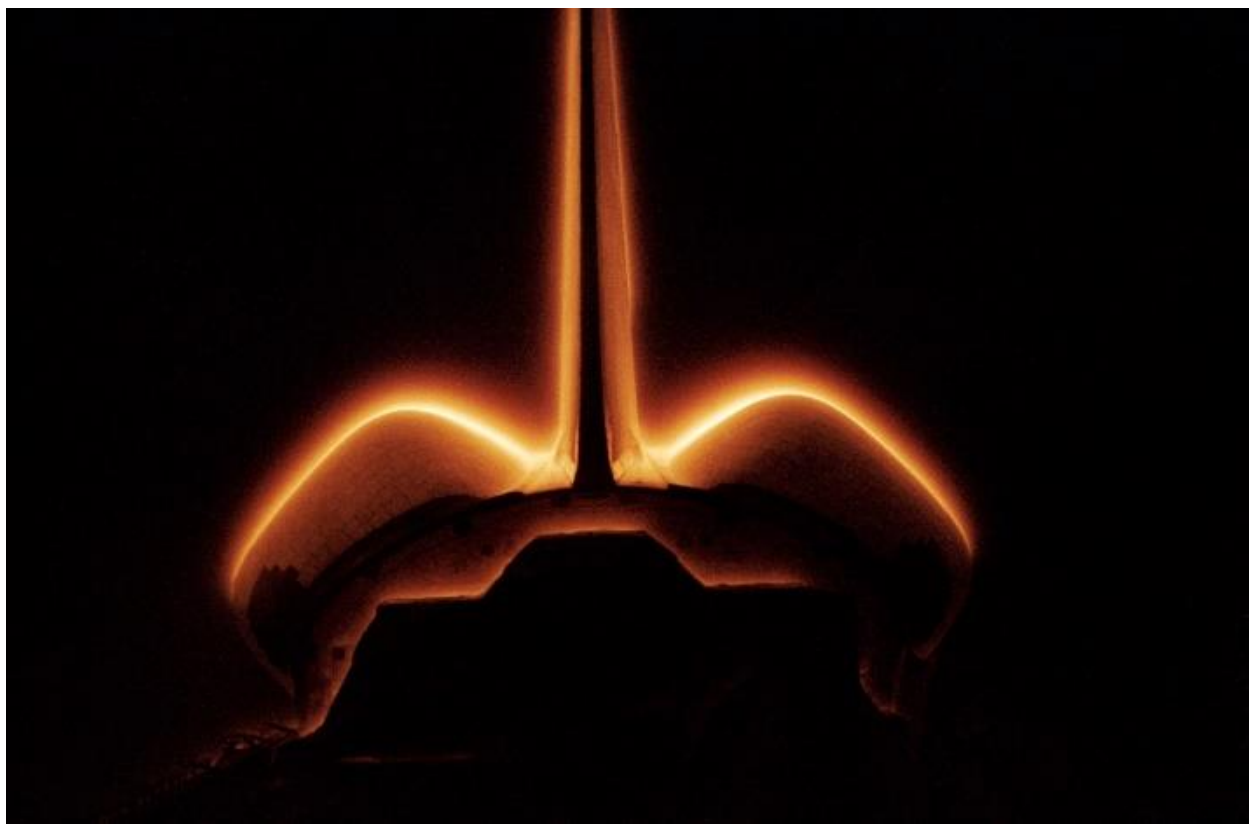

Figure 11 – Plasma glowing on Shuttle Columbia during the STS-62 (1994). Credits: NASA

A positive electrode immersed in plasma will collect electrons and repel ions. This leads to the formation of a surrounding volume influenced by the charged electrode and called "sheath". A measure of its size is given by the Debye length and electrons outside it will not experience any perturbation given by the electrode itself:

$$\lambda_D = \sqrt{\frac{K \cdot T_e}{4\pi e^2 n}}$$

where:

K is the Boltzmann's constant;

$T_e$ is the electron temperature (K);

e is the electron charge;

n is the plasma density in the surrounding environment.

Generally, the sheath will not be thick than 10 Debye lengths.

The effect of the plasma on spacecraft are not the same as ionizing radiations, yet can induce a charge on the spacecraft due to the flux of positive and negative ions so that differences in potential can be thousands of volts between different components **(48)** and result in electric arcing or biasing of instrument readings. These problems can eventually lead to upsets to electronics and possible failures.

It must be noted how Albarran et.al. **(49)** highlights that spacecraft–plasma interactions (surface charging, plasma sheaths, and wakes) are less understood for CubeSats as their dimensions better match the Debye length of the plasma environment in LEO.

### *6.2 Countermeasures*

The first attempt against spacecraft charging comes again from the MISSE experiment **(39)** in which Nasa's Electrodynamic Dust Shield (EDS) was tested onboard the ISS **(45)**. Originally intended to avoid suit charging due to Martian and Lunar regolith, the EDS has tested with Indium Tin Oxide (ITO) electrodes on glass, along with either a polyethylene terephthalate (PET) or a broadband anti-reflective (BBAR) coatings.

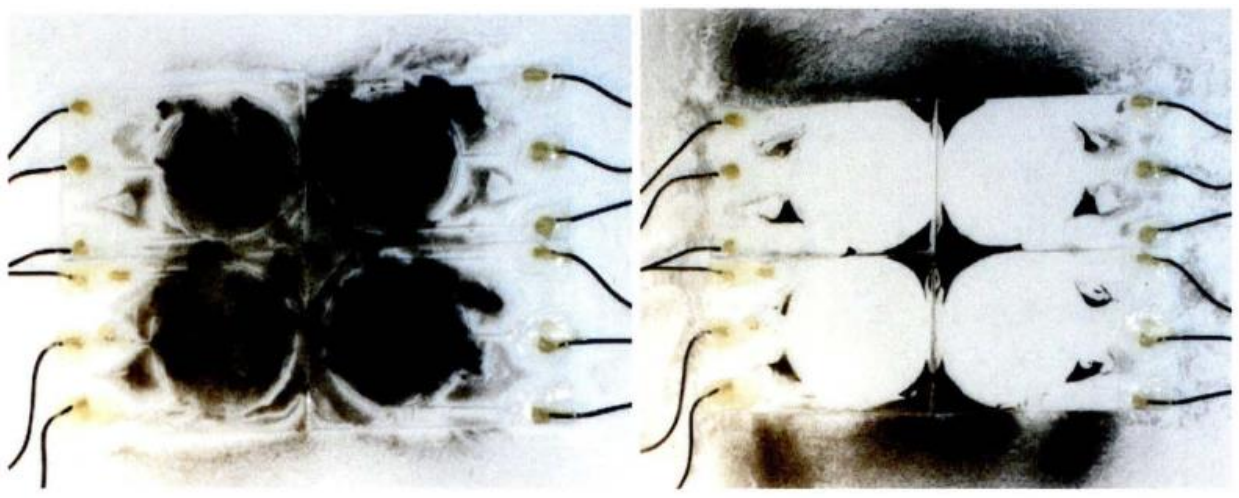

Figure 12 - Left: dust deposition and EDS off. Right: dust cleaned by turning the EDS on **(50)**.

In the first case, the EDS went from -2090 V down to -41,4 V in 70,83 hours, about 47 ISS orbits which were indeed unacceptable. In the case of BBAR, the EDS passed from -1930 V to -15,3 V in 21,67 minutes, clearly representing a better choice, also considering that it meets the requirements of MIL-C-675C, MIL-C-14806A, and MIL-C-48497A.

Despite the good results, this application might not be fit for small satellites, given the additional power requirement and the possible electric interaction with adjacent components.

General recommendations contained in the NASA-HDBK-4006A also call into question controlling the spacecraft potential by forcing the spacecraft to the ambient plasma potential ("plasma contactor solution"), preventing the exposure to the plasma of any high-voltage conducting surface (encapsulation), electrically bonding different surfaces to prevent any internal difference in potential (addressed in MSFC-HDBK-3697, Electrical Bonding Design Guide Handbook, and NASA-STD-4003, Electrical Bonding for NASA Launch Vehicles, Spacecraft, Payloads, and Flight Equipment). A thorough discussion can be found in Ch. 7 of the NASA-HDBK-4006A.

All these solutions are, again, originally intended for large spacecrafts and possibly the ISS itself, but very little is found about the growing world of nanosat. A possible solution was proposed for Nasa's Shield-1 nanosat **(35)** and represented by a Charge Dissipation Film Resistance called LUNA XP-CD-B which is characterized by a resistivity in the order of $10^8$-$10^{10}$ ohm-cm.

For this on-orbit demonstration, as for figure 14, the film was set between two electrodes and connecter to a fixed current source for a 2-wire resistance measurement over the time with an expected resistance of 2 MOhm at 25 °C using a 5 $cm^2$ sample 25 µm thick (its typical coating thickness). The electronics of the nanosat itself are bound to be put in a LUNA XP-CD-B coated vault to further decrease the risk of internal charging.

# APPENDIX A

Table 3. Atomic Oxygen (AO) Erosion Yields $E_y$ From MISSE 2–8 Ram Polymer Experiments

| Material [trade names][a] | Sample abbreviation | Thickness, mil | MISSE 2 ram AO = $8.43\times10^{21}$ atom/cm$^2$ Solar = 6,300 ESH Mission: 4 years (Ref. 18) | | MISSE 4 ram AO = $2.15\times10^{21}$ atom/cm$^2$ Solar = 1,400 ESH Mission: 1 year (Ref. 28) | | MISSE 6B ram (nonstressed) AO = $1.97\times10^{21}$ atom/cm$^2$ Solar = 2,600 ESH Mission: 1.5 years (Ref. 19) | | MISSE 7B ram AO = $4.22\times10^{21}$ atom/cm$^2$ Solar = 2,400 ESH Mission: 1.5 years (Ref. 21) | | MISSE 8 ram AO = $4.62\times10^{21}$ atom/cm$^2$ Solar ≈ 3,200 ESH Mission: 2 years (Ref. 22) | |
|---|---|---|---|---|---|---|---|---|---|---|---|---|
| | | | ID (layers) (Ref. 2) | AO $E_y$, cm$^3$/atom | ID (layers) | AO $E_y$, cm$^3$/atom | ID (layers) | AO $E_y$, cm$^3$/atom | ID[b] (layers) | AO $E_y$, cm$^3$/atom | ID (layers) | AO $E_y$, cm$^3$/atom |
| Amorphous fluoropolymer [Teflon® AF 1601] | AF | 2 | 2-E5-45 (1) | $1.98\times10^{-25}$ | | | | | | | | |
| Carbon: pyrolytic graphite, C plane | PG | 80 | 2-E5-25 (1) | $4.15\times10^{-25}$ | | | | | | | | |
| Crystalline polyvinyl fluoride with white pigment [White Tedlar®] | PVF-W | 2 | | | | | | | B7-1 (1) | $1.48\times10^{-25}$ | | |
| Crystalline polyvinyl fluoride with white pigment and a 0.5-mil Kapton® H cover [White Tedlar®] | PVF-W | 2 | | | | | | | B7-2 (1) AO fluence = $3.79\times10^{21}$ | $1.54\times10^{-25}$ | | |
| Crystalline polyvinyl fluoride with white pigment and two 0.5-mil Kapton® H covers [White Tedlar®] | PVF-W | 2 | | | | | | | B7-3 (1): AO fluence = $3.37\times10^{21}$ | $1.67\times10^{-25}$ | | |
| Crystalline polyvinyl fluoride with white pigment [White Tedlar® TW10B53] | PVF-W | 1 | 2-E5-11 (13) | $1.01\times10^{-25}$ | | | | | | | M8-R5 (7) | $1.45\times10^{-25}$ |
| Hubble Space Telescope fluorinated ethylene propylene/Al, 1st servicing mission, SM1 MSS-D: 3.6 yr, 11,339 ESH | HST FEP/Al | 5 | | | | | W6-8 (1)[f] | $2.40\times10^{-25}$ | | | | |
| Hubble Space Telescope fluorinated ethylene propylene/Al, 3rd servicing | HST FEP/Al | 5 | | | | | W6-7 (1)[f] | $2.11\times10^{-25}$ | | | M8-R6 (1) | $2.50\times10^{-25}$ |
| Polyimide (PMDA) [Kapton® H] | Kapton H | 5 | 2-E5-30 (3), 2-E5-33 (3) | [h]$3.00\times10^{-24}$ | 2-E22-18 (7) | [h]$3.00\times10^{-24}$ | W2-3/N8 (4), W6-5/GW-1 (3) | [h]$3.00\times10^{-24}$ | B7-8 (4) | [h]$3.00\times10^{-24}$ | M8-R1 (4) | [h]$3.00\times10^{-24}$ |
| Polyimide, clear [CP1] | CP1 | 1 | 2-E6-15 (9)[e] | $1.91\times10^{-24}$ | 2-E22-26 (20) | $2.26\times10^{-24}$ | W2-4/N4 (20) | $2.16\times10^{-24}$ | | | | |
| Polyimide (BPDA) [Upilex®-S] | PI (Upilex-S) | 1 | 2-E5-32 (11) | $9.22\times10^{-25}$ | 2-E22-19 (21) | $1.71\times10^{-24}$ | W2-15/N3 (20) | $1.65\times10^{-24}$ | | | | |
| Silicone [DC 93-500 on fused silica] | DC 93-500 | 10 | | | | | | | | | M8-R10 (1) | $3.81\times10^{-27}$ |
| Triton oxygen resistant, low modulus polymer[i] [TOR™ LM] | TOR | 1.5 | | | 2-E22-30 (1) | Mass gain | | | | | | |
| Urethane/Vectran™ mesh | Ur/Vectra | 10 | | | | | | | N5-16 (1) | $4.61\times10^{-25}$ | | |